\documentclass[conference,a4paper]{IEEEtran}

\usepackage{amsmath}
\usepackage{amssymb}
\usepackage{latexsym}
\usepackage{array,arydshln}
\usepackage{multirow}
\usepackage{graphicx}
\usepackage{float}
\usepackage{bm}
\usepackage{breqn}
\usepackage{ctable}

\newcommand{\otoprule}{\midrule[\heavyrulewidth]}
\newcommand{\bs}[1]{\ensuremath{\boldsymbol{#1}}}

\renewcommand{\u}{\bs{u}}
\newcommand{\CU}{C_{\mathrm{U}}}
\newcommand{\CL}{C_{\mathrm{L}}}
\newcommand{\CO}{C_{\mathrm{O}}}
\newcommand{\CI}{C_{\mathrm{I}}}
\newcommand{\DU}{D_{\mathrm{U}}}
\newcommand{\DL}{D_{\mathrm{L}}}

\newcommand{\CPCCa}{\mathcal{C}_{\mathrm{PCC}}}
 
\newcommand{\CPCCSa}{\mathcal{C}_{\mathrm{SC-PCC}}}

\newcommand{\CSCCa}{\mathcal{C}_{\mathrm{SCC}}}
 
\newcommand{\CSCCSa}{\mathcal{C}_{\mathrm{SC-SCC}}}

\newcommand{\ut}[2]{\ensuremath{\bs{u}_{#1,\mathrm{#2}}}}

\newcommand{\fs}[1]{\ensuremath{f_{\mathrm{#1},\mathrm{s}}}}
\newcommand{\fp}[1]{\ensuremath{f_{\mathrm{#1},\mathrm{p}}}}

\newcommand{\xsa}[2]{\ensuremath{p_{\mathrm{#1},\mathrm{#2}}}}
\newcommand{\ysa}[2]{\ensuremath{p_{\mathrm{#1},\mathrm{#2}}}}

\newcommand{\xs}[4]{\ensuremath{p_{\mathrm{#1},\mathrm{#2}}^{(#3,#4)}}}

\IEEEoverridecommandlockouts

\begin{document}

\title{Spatially Coupled Turbo Codes}


\author{
\IEEEauthorblockN{Saeedeh Moloudi$^\dag$, Michael Lentmaier$^\dag$, and Alexandre Graell i Amat$^\ddag$}
\IEEEauthorblockA{$\dag$Department of Electrical and Information
  Technology, Lund University, Lund, Sweden \\
  $\ddag$Department of Signals and Systems, Chalmers University of Technology, Gothenburg, Sweden\\
              \{saeedeh.moloudi,michael.lentmaier\}@eit.lth.se, alexandre.graell@chalmers.se}\\
              \thanks{This work was supported in part by the Swedish Research Council (VR) under grant \#621-2013-5477.}\vspace*{-1cm}
}


\maketitle

\begin{abstract}
In this paper, we introduce the concept of spatially coupled turbo codes (SC-TCs), as the turbo codes counterpart of spatially coupled low-density parity-check codes. We describe spatial coupling for both Berrou \textit{et al.} and Benedetto \textit{et al.} parallel and serially concatenated codes. For the binary erasure channel, we derive the exact density evolution (DE) equations of SC-TCs by using the method proposed by Kurkoski \textit{et al.} to compute the decoding erasure probability of convolutional encoders. Using DE, we then analyze the asymptotic behavior of SC-TCs. We observe that the belief propagation (BP) threshold of SC-TCs improves with respect to that of the uncoupled ensemble and approaches its maximum a posteriori threshold. This phenomenon is especially significant for serially concatenated codes, whose uncoupled ensemble suffers from a poor BP threshold.
\end{abstract}

\IEEEpeerreviewmaketitle

\section{Introduction}


Low-density parity-check (LDPC) convolutional codes \cite{JimenezLDPCCC}, also known as spatially coupled LDPC (SC-LDPC) codes \cite{Kudekar_ThresholdSaturation}, can be obtained from a sequence of individual LDPC block codes by distributing the edges of their Tanner graphs over several adjacent blocks \cite{LentmaierTransITOct2010}. The resulting spatially coupled  codes exhibit a threshold saturation phenomenon, which has attracted a lot of interest in the past few years: the threshold of an iterative belief propagation (BP) decoder, obtained by density evolution (DE), is improved to that of the optimal maximum-a-posteriori (MAP) decoder \cite{Kudekar_ThresholdSaturation,LentmaierTransITOct2010}. As a consequence, it is possible to achieve capacity with simple regular LDPC codes, which show without spatial coupling a significant gap between BP and MAP threshold. 

The concept of spatial coupling is not limited to LDPC codes. Spatially coupled turbo-like codes, for example, can be obtained by replacing the block-wise permutation of a turbo code by a convolutional permutation \cite{ldcc_2}. In combination with a windowed decoder for the component code, a continuous streaming implementation is possible \cite{Hall2001}. The self-concatenated convolutional codes in \cite{AAECC} are closely related structures as well. A variant of spatially coupled self-concatenated codes with block-wise processing, called 
laminated codes was considered in \cite{Huebner2008}. They have the
advantage that an implementation similar to uncoupled turbo codes is possible, without the need for
a streaming implementation of the decoder. A block-wise version of braided convolutional codes \cite{ZhangBCC}, a class of spatially coupled codes with convolutional components, has recently been analyzed in \cite{MoloudiISIT14}. 

The aim of this paper is to investigate the impact of spatial coupling
on the thresholds of \textit{classical} turbo codes. For this purpose we introduce some
special block-wise spatially coupled ensembles of parallel concatenated codes (SC-PCCs)
and serially concatenated codes (SC-SCCs), which are spatially coupled
versions of the ensembles by Berrou {\em et al.}  \cite{BerrouTC}  and
Benedetto {\em et al.} \cite{Benedetto98Serial}, respectively. With a slight abuse of the term, we call both parallel and serial ensembles spatially coupled turbo codes (SC-TCs). For these ensembles we derive exact DE equations from the transfer functions of the component decoders \cite{Kur03,tenBrinkEXITConv} and perform a threshold analysis for the binary erasure channel (BEC), analogously to \cite{LentmaierTransITOct2010,MoloudiISIT14}. To compare the results for SC-PCC and SC-SCC ensembles with each other some ensembles with puncturing are also considered. The BP thresholds of the different ensembles are presented and compared to the MAP thresholds for different coupling memories.

\section{Spatially Coupled Turbo Codes}
\label{secEnsembles}

In this section, we introduce spatially coupled turbo codes. We first describe spatial coupling for both parallel and serially concatenated codes, and then address their iterative decoding.
 
\subsection{Spatially Coupled Parallel Concatenated Codes}
 
\begin{figure*}[!t]
  \centering
    \includegraphics[width=1\textwidth]{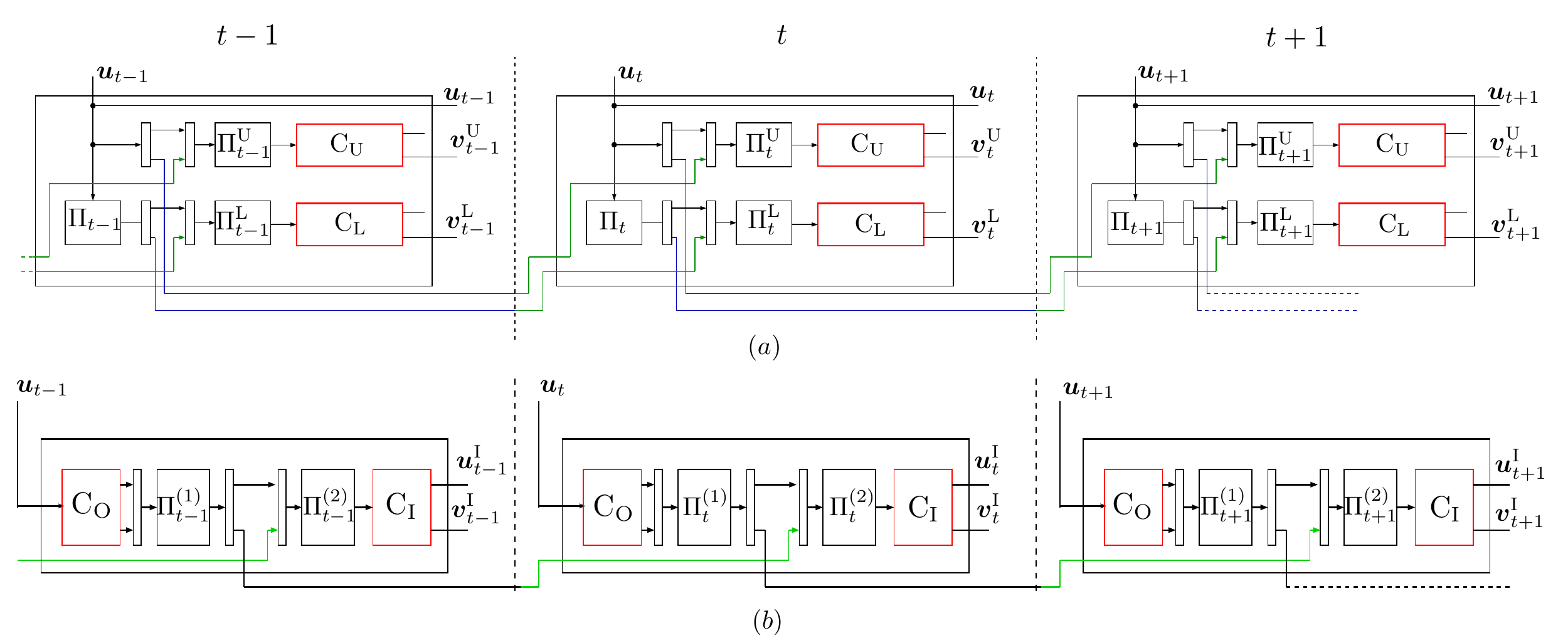}
    \vspace{-4ex}
\caption{Block diagram of the encoder of a spatially coupled turbo code for coupling memory $m=1$. (a) parallel concatenation (b) serial concatenation.}
\label{Encoder}
\vspace{-2ex}
\end{figure*}

We consider the spatial coupling of $R = 1/3$ parallel concatenated codes, built from the parallel concatenation of two rate-$1/2$ recursive systematic convolutional encoders, denoted by
$\CU$ and $\CL$ (see Fig.~\ref{Encoder}). For simplicity, we describe spatial coupling with coupling memory $m=1$. Consider a collection of $L$ turbo encoders at time instants $t=1,\ldots,L$, as illustrated in Fig.~\ref{Encoder}(a). $L$ is called the coupling length. We denote by $\bs{u}_t$ the information sequence, and by
$\bs{v}_t^{\text{U}}$ and $\bs{v}_t^{\text{L}}$ the code sequences of
$C_{\text{U}}$  and $C_{\text{L}}$, respectively, at time $t$. The
output of the turbo encoder is given by the tuple $\bs{v}_t =
(\bs{u}_t,\bs{v}^{\text{U}}_t ,\bs{v}^{\text{L}}_t )$. A SC-PCC
ensemble (with $m=1$) is obtained by connecting each turbo code in
the chain to the one on the left and to the one on the right as
follows. Divide the information sequence $\u_t$ into two
sequences, $\ut{t}{A}$ and $\ut{t}{B}$ by a demultiplexer. Also divide
a copy of the information sequence, which is properly reordered by the permutation $\Pi_{t}$, into two
sequences, $\ut{t}{A'}$ and $\ut{t}{B'}$ by another demultiplexer. At time $t$, the information sequence at the input of encoder $\CU$ is $(\ut{t}{A},\ut{t-1}{B})$, properly reordered by a permutation $\Pi^{\text{U}}_t$. Likewise, the information sequence at the input of encoder $\CL$ is $(\ut{t}{A'},\ut{t-1}{B'})$, properly reordered by the permutation $\Pi^{\text{L}}_t$. 
In Fig.~\ref{Encoder} the blue lines represent the information bits
from the current time slot $t$ that are used in the next time slot $t+1$ and the green lines represent the
information bits from the previous time slot $t-1$. 
In order to terminate the encoder of the SC-PCC to the zero state, the information
sequences at the end of the chain are chosen in such a way that the
output sequence at time $t=L+1$ becomes
$\bs{v}_{L+1}=\bs{0}$. Analogously to conventional convolutional codes
this results in a rate loss that becomes smaller as $L$ increases.  

Using the procedure described above a coupled chain (a convolutional
structure over time) of $L$ turbo encoders with coupling memory
$m=1$ is obtained. An extension to larger coupling memories
$m>1$ is presented in Section~\ref{secMem}.

\subsection{Spatially Coupled Serially Concatenated Codes}

We consider the coupling of serially concatenated codes (SCCs) built from the serial concatenation of two rate-$1/2$ recursive systematic convolutional encoders. The overall code rate of the uncoupled ensemble is therefore $R=1/4$. A block diagram of the encoder is depicted in Fig.~\ref{Encoder}(b) for coupling memory $m=1$. As for SC-PCCs, let $\bs{u}_t$ be the information sequence at time $t$. Also, denote by $\bs{v}_t^{\mathrm{O}}=(\bs{v}_t^{\mathrm{O,s}},\bs{v}_t^{\mathrm{O,p}})=(\bs{u}_t,\bs{v}_t^{\mathrm{O,p}})$ and $\bs{v}_t^{\mathrm{I}}$ the encoded sequence at the output of the outer and inner encoder, respectively, and by $\tilde{\bs{v}}_t^{\mathrm{O}}$ the sequence $\bs{v}_t^{\mathrm{O}}$ after permutation. The SC-SCC with $m=1$ is constructed as follows. Consider a collection of $L$ SCCs at time instants $t=1,\ldots,L$. Divide the sequence $\tilde{\bs{v}}_t^{\mathrm{O}}$ into two parts, $\tilde{\bs{v}}_{t,\text{A}}^{\mathrm{O}}$ and
$\tilde{\bs{v}}_{t,\text{B}}^{\mathrm{O}}$. Then, at time $t$ the
sequence at the input of the inner encoder $\CI$ is
$(\tilde{\bs{v}}_{t,\text{A}}^{\mathrm{O}},\tilde{\bs{v}}_{t-1,\text{B}}^{\mathrm{O}})$. 
In order to terminate the encoder of the SC-SCC to the zero state, the information
sequences at the end of the chain are chosen in such a way that the
output sequence at time $t=L+1$ becomes
$\bs{v}_{L+1}^{\text{I}}=\bs{0}$.

Using this construction method, a coupled chain of $L$ SCCs with coupling memory $m=1$ is obtained. An extension to larger coupling memories
$m>1$ is presented in Section~\ref{secMem}.

\subsection{Iterative decoding}

\begin{figure}[!t]
  \centering
    \includegraphics[width=\linewidth]{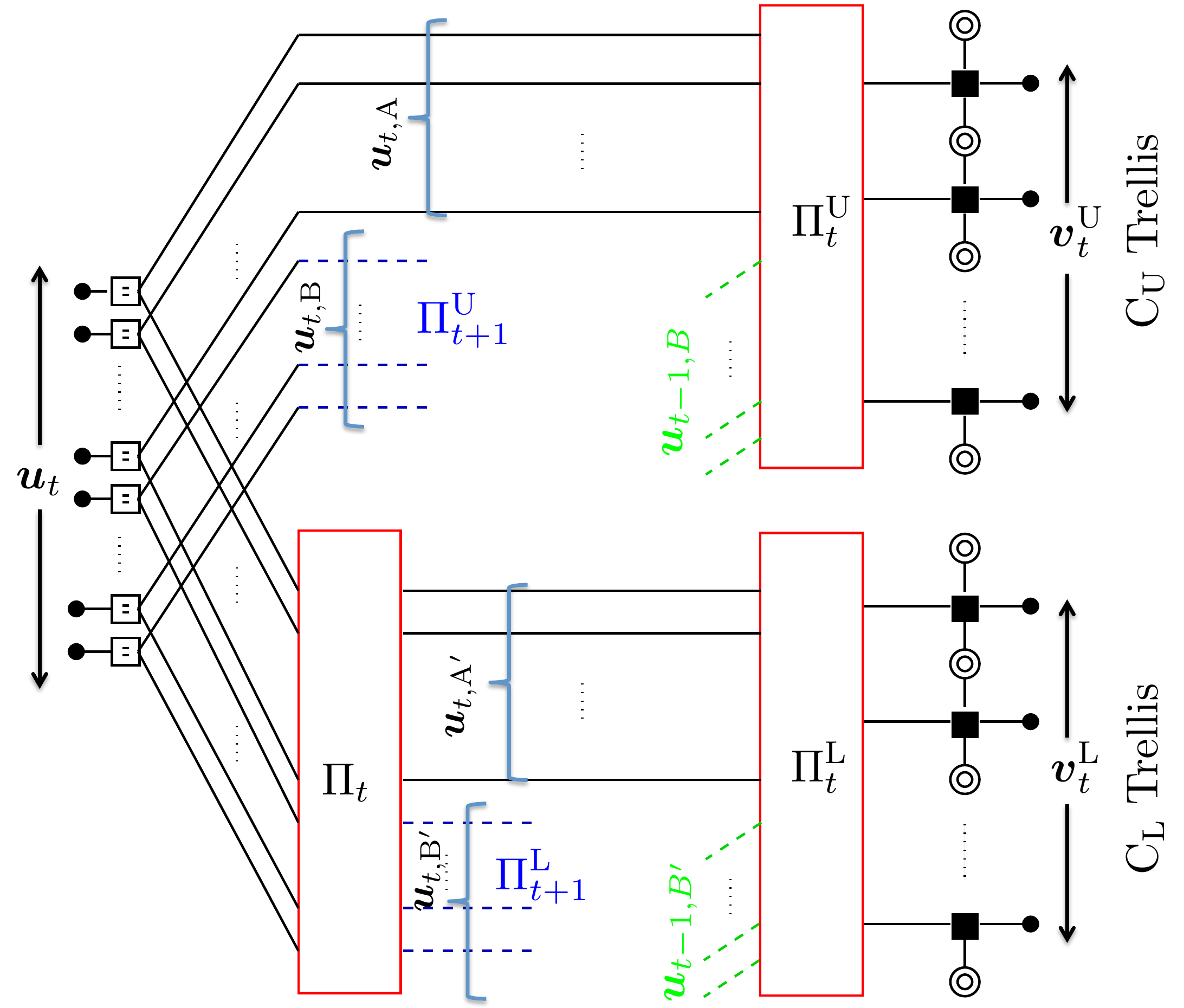}
\caption{Factor graph of a single section of a SC-PCC.}
\label{factorP}
\vspace{-2ex}
\end{figure}

As standard turbo codes, SC-TCs can be decoded using iterative message passing (belief propagation) decoding, where the component encoders of each turbo code are decoded using the BCJR algorithm. The BP decoding of SC-PCCs can be easily visualized with the help of Fig.~\ref{factorP}, which shows the factor graph of a single section of the SC-PCC. We denote by $\DU$ and $\DL$ the decoder of the upper and lower encoder, respectively.

The decoder $\DU$ receives at its input information from the channel for both systematic and parity bits. Furthermore, it also receives a-priori information on the systematic bits from
other decoders. As described above, at time $t$ the information
sequence at the input of $\CU$ consists of two parts,
$\bs{u}_{t,\text{A}}$ and $\bs{u}_{t-1,\text{B}}$. Correspondingly,
$\DU$ at time instant $t$ receives a priori information from $\DL$ at time instants $t-1$,
$t$ and $t+1$. Based on the information from the channel and from the companion decoders, $\DU$ computes the extrinsic information on the systematic bits using the BCJR algorithm. Since the structure of SC-PCCs is symmetric, the decoding of the lower encoder is performed in an identical manner.

\begin{figure}[!t]
  \centering
    \includegraphics[width=.9\linewidth]{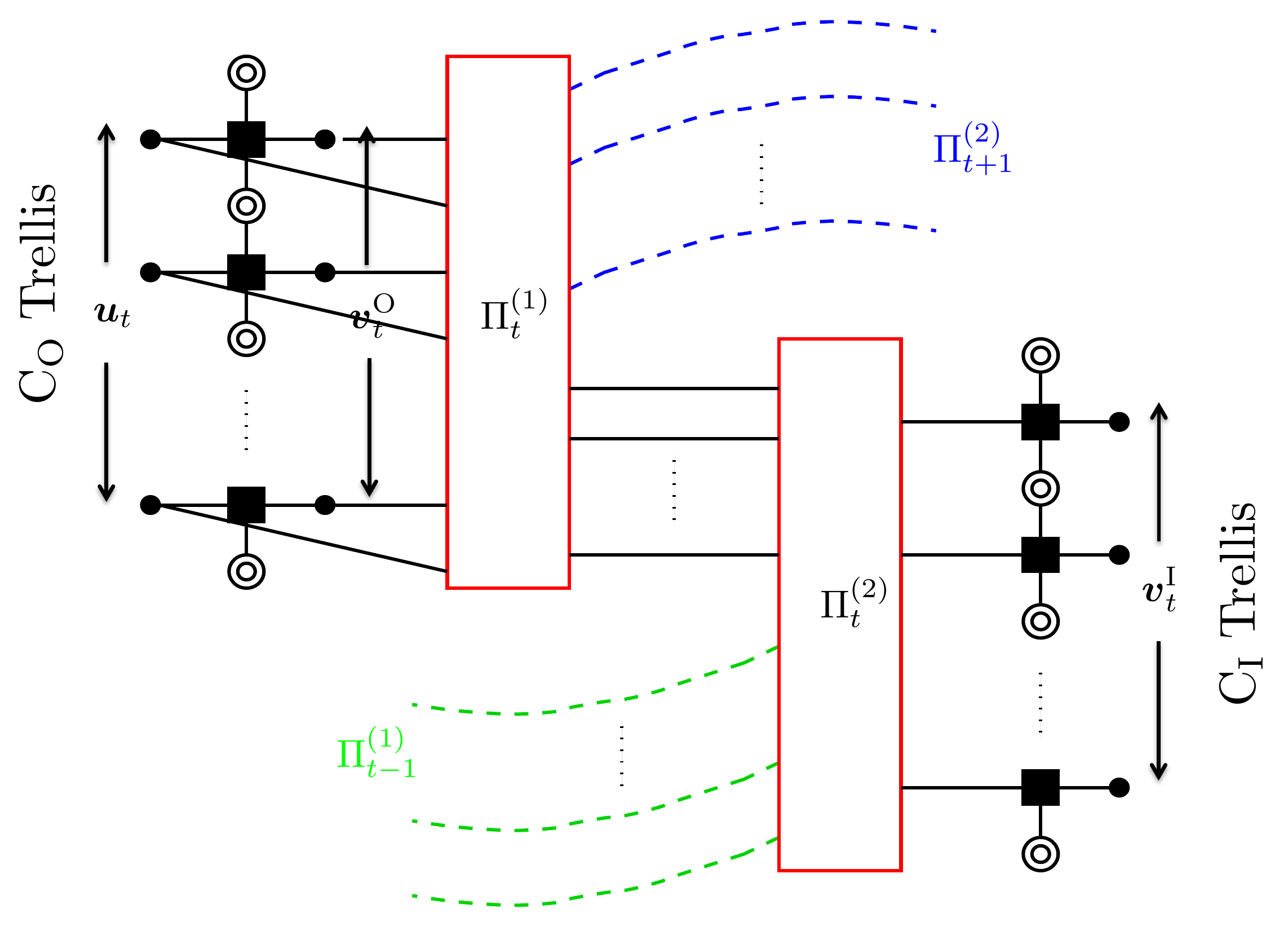}
\caption{Factor graph of a single section of a SC-SCC}
\label{factorS}
\vspace{-2ex}
\end{figure}

Similarly to SC-PCCs, the decoding SC-SCCs can also be described with the help of a factor graph. The factor graph of a section of a SC-SCC with $m=1$ is shown in Fig.~\ref{factorS}. 

\section{Density Evolution Analysis on the BEC}
\label{secDE}

In this section, we analyze the asymptotic performance of SC-TCs using DE. We consider transmission over a BEC with erasure probability $\epsilon$, denoted by BEC$(\epsilon)$. We derive the exact DE equations for both (unpunctured) SC-PCCs and SC-SCCs and discuss the modification of the equations when puncturing is applied for achieving higher rates.

\subsection{Spatially Coupled Parallel Concatenated Codes} 

Let $\xsa{U}{s}$ and $\ysa{L}{s}$ be the average (extrinsic) erasure probability on the systematic bits at the output of the upper and lower decoder, respectively. Likewise, we define $\xsa{U}{p}$ and $\ysa{L}{p}$ for the parity bits.

The erasure probabilities $\xsa{U}{s}$ and $\xsa{U}{p}$ at iteration $i$ and time instant $t$ can be written as
\begin{align}
\label{eq:UpperUpdates}
\xs{U}{s}{i}{t}&=\fs{U}\left(
q_{\text{L}}^{(i-1)},\epsilon\right)\\
\label{eq:UpperUpdatep}
\xs{U}{p}{i}{t}&=\fp{U}\left(
q_{\text{L}}^{(i-1)},\epsilon\right),
\end{align}
where
\begin{equation}
\label{eq:UpperUpdate3}
q_{\text{L}}^{(i-1)}=\epsilon \cdot\frac{2\xs{L}{s}{i-1}{t}+\xs{L}{s}{i-1}{t-1}+\xs{L}{s}{i-1}{t+1}}{4},
\end{equation}
and $\fs{U}$ and $\fp{U}$ denote the upper decoder transfer functions for the systematic and parity bits, respectively. 

Note that the upper decoder transfer function at time $t$ depends on
both the channel erasure probability and the extrinsic erasure
probability on the systematic bits from the lower decoder at time
instants $t$,  $t-1$ and $t+1$, due to the coupling. Because of the symmetric design, the lower decoder update is identical to that of the upper decoder by interchanging $p_{\rm{U}}$ and $p_{\rm{L}}$, and substituting $q_{\rm{L}}\leftarrow q_{\rm{U}}$ in (\ref{eq:UpperUpdates})--(\ref{eq:UpperUpdate3}).

Finally, the a-posteriori erasure probability on the information bits at time $t$ and iteration $i$ is\footnote{We remark that although \eqref{eq:UpperUpdatep} is not applied within the DE recursion it is required for the computation of the area bound on the MAP threshold.}
 \begin{equation}
\label{eq:appPCC}
\resizebox{.9\hsize}{!}{$
p^{(i,t)}_{\rm a}=\epsilon\cdot
\frac{\xs{U}{s}{i}{t}\xs{L}{s}{i}{t}+\xs{U}{s}{i}{t}\xs{L}{s}{i}{t+1}+\xs{U}{s}{i}{t+1}\xs{L}{s}{i}{t}+\xs{U}{s}{i}{t+1}\xs{L}{s}{i}{t+1}}{4}$}
\end{equation}

For the BEC it is possible to compute analytic expressions for the
exact (extrinsic) probability of erasure of convolutional encoders,
using the method proposed in \cite{Kur03} and
\cite{tenBrinkEXITConv}. Here, we use this method to derive the exact
expressions for the transfer functions of the component decoders. DE is then performed by
tracking the evolution of $p^{(i,t)}_{\rm a}$ with the number of
iterations, with the initialization
$\xs{U}{s}{0}{t}=\xs{U}{p}{0}{t}=\xs{L}{s}{0}{t}=\xs{L}{p}{0}{t}=0$
for $t=0$ and $t>L$, and 1 otherwise. The BP threshold corresponds to
the maximum channel parameter $\epsilon$ for which successful decoding
is achieved, i.e.,  $p^{(i,t)}_{\rm a}$ tends to zero  for all time
instants $t$  as $i$ tends to infinity.

\subsection{Spatially Coupled Serially Concatenated Codes}

Similarly to the parallel case, DE equations can be derived for SC-SCCs. Let $\xsa{O}{s}$ and $\ysa{I}{s}$ be the average (extrinsic) erasure probability on the systematic bits at the output of the outer and inner decoder, respectively. Likewise, we define $\xsa{O}{p}$ and $\xsa{I}{p}$ for the parity bits at the output of the outer and inner decoder, respectively.

The erasure probabilities $\ysa{I}{s}$ and $\ysa{I}{p}$ can be written as
\begin{align}
\label{eq:LowerUpdates}
\xs{I}{s}{i}{t}&=\fs{I}\left(
q_{\text{O}}^{(i-1)},\epsilon\right)\\
\label{eq:LowerUpdatep}
\xs{I}{p}{i}{t}&=\fp{I}\left(
q_{\text{O}}^{(i-1)},\epsilon\right),
\end{align}
where
\begin{equation}
\label{eq:InnerUpdate3}
q_{\text{O}}^{(i-1)}=\epsilon \cdot\frac{\xs{O}{s}{i-1}{t}+\xs{O}{p}{i-1}{t}+\xs{O}{s}{i-1}{t-1}+\xs{O}{p}{i-1}{t-1}}{4},
\end{equation}
and $\fs{I}$ and $\fp{I}$ denote the inner decoder transfer functions for the systematic and parity bits, respectively. 

Likewise, $\ysa{O}{s}$ and $\ysa{O}{p}$ are
\begin{align}
\label{eq:OuterUpdates}
\xs{O}{s}{i}{t}&=\fs{O}\left(q_{\text{I}}^{(i-1)},q_{\text{I}}^{(i-1)}\right)\\
\label{eq:OuterUpdatep}
\xs{O}{p}{i}{t}&=\fp{O}\left(q_{\text{I}}^{(i-1)},q_{\text{I}}^{(i-1)}\right),
\end{align}
where
\begin{equation}
\label{eq:OuterUpdate3}
q_{\text{I}}^{(i-1)}=\epsilon \cdot \frac{\xs{I}{s}{i-1}{t}+\xs{I}{s}{i-1}{t+1}}{2} \ .
\end{equation}

The a-posteriori erasure probability on the information bits at time $t$ after
$i$ iterations is
\begin{equation}
\label{eq:appSCC}
p^{(i,t)}_{\rm a}=\epsilon \cdot \xs{O}{s}{i}{t} \cdot \frac{\xs{I}{s}{i}{t}+\xs{I}{s}{i}{t+1}}{2} \ .
\end{equation}

DE is then performed by tracking the evolution of $p^{(i,t)}_{\rm a}$ with the number of iterations, with the initialization $\xs{I}{s}{0}{t}=\xs{I}{p}{0}{t}=\xs{O}{s}{0}{t}=\xs{O}{p}{0}{t}=0$ for $t=0$ and $t>L$ and 1 otherwise.

\subsection{Spatially Coupled Turbo Codes with Random Puncturing}

Higher rates can be obtained by applying puncturing. Here, we consider random puncturing. Assume that a code sequence $\bs{x}$ is randomly punctured such that a fraction $\rho\in[0,1]$ of the coded bits survive after puncturing, and then transmitted over a BEC$(\epsilon)$. $\rho$ will be referred to as the \textit{permeability rate}. For the BEC, puncturing is equivalent to transmitting $\bs{x}$ through a BEC$(\epsilon_{\rho})$ resulting from the concatenation of two BECs, BEC$(\epsilon)$ and BEC$(1-\rho)$, where $\epsilon_{\rho}=1-(1-\epsilon)\rho$. The DE equations derived in the previous subsections can be easily modified to account for puncturing. Consider first the case of SC-PCCs. We consider only puncturing of the parity bits, and that both $\CU$ and $\CL$ are equally punctured with permeability rate $\rho$. The code rate of the (uncoupled) punctured parallel concatenated code (PCC) is $R=\frac{1}{1+2\rho}$. This results in a slight modification of the DE equations, substituting $\epsilon\leftarrow\epsilon_{\rho}$ in (\ref{eq:UpperUpdates}), (\ref{eq:UpperUpdatep}).

For SC-SCCs we consider puncturing as proposed in \cite{AGiAa,AGiAb}, which results in better SCCs as compared to standard SCCs. Let $\rho_0$ and $\rho_1$ be the permeability rate of the systematic and parity bits, respectively, of $\CO$ sent directly to the channel (see \cite[Fig.~1]{AGiAb}), and $\rho_2$ the permeability rate of the parity bits of $\CI$. The code rate of the (uncoupled) punctured\footnote{In this paper we consider $\rho_0=1$, i.e., the overall code is systematic.} SCC is $R=\frac{1}{\rho_0+\rho_1+2\rho_2}$. The DE for punctured SC-SCCs is obtained by substituting $\epsilon\leftarrow\epsilon_{\rho_2}$ in (\ref{eq:LowerUpdates}), (\ref{eq:LowerUpdatep}), and modifying (\ref{eq:InnerUpdate3}) to
\begin{align*}
&q_{\text{O}}^{(i-1)}=\\
& \frac{\epsilon\cdot\left(\xs{O}{s}{i-1}{t}+\xs{O}{s}{i-1}{t-1}\right)+\epsilon_{\rho_1}\cdot\left(\xs{O}{p}{i-1}{t}+\xs{O}{p}{i-1}{t-1}\right)}{4},
\end{align*}
and (\ref{eq:OuterUpdates}), (\ref{eq:OuterUpdatep}) to
\begin{align}
\label{eq:OuterUpdatesPunct}
\xs{O}{s}{i}{t}&=\fs{O}\left(q_{\text{I}}^{(i-1)},{\tilde{q}}_{\text{I}}^{(i-1)}\right)\\
\label{eq:OuterUpdatepPunct}
\xs{O}{p}{i}{t}&=\fp{O}\left(q_{\text{I}}^{(i-1)},{\tilde{q}}_{\text{I}}^{(i-1)}\right),
\end{align}
where $q_{\text{I}}^{(i-1)}$ is given in (\ref{eq:OuterUpdate3}) and
\begin{equation}
\label{eq:OuterUpdate3Punct}
\tilde{q}_{\text{I}}^{(i-1)}=\epsilon_{\rho_1} \cdot \frac{\xs{I}{s}{i-1}{t}+\xs{I}{s}{i-1}{t+1}}{2} \ .
\end{equation}

\section{Extension to Larger Coupling Memories}
\label{secMem}

The results from the previous sections can easily be generalized to larger coupling memories $m >1$. 

Let us first consider SC-PCCs. In the general case the information
sequences $\bs{u}_{t}, \bs{u}_{t-1}, \dots, \bs{u}_{t-m}$ from
$m+1$ different time instances are used by the encoders at time
$t$.  This is achieved by  dividing the  information sequence
$\bs{u}_t$ into the sequences $\bs{u}_{t,j}$,
$j=0,\dots,m$ by a multiplexer, and also dividing a properly
reordered copy of the information
bits into $\bs{u}_{t,j'}$,
$j'=0,\dots,m$, which can be accomplished by permutation $\Pi_t$
followed by a multiplexer. At the input of the upper encoder
$\CU$ at time $t$ the sequences $\bs{u}_{t-j,j}$ are multiplexed and
reordered by the permutation $\Pi_{t}^{\text{U}}$. The lower encoder
$\CL$ receives the information sequences
$\bs{u}_{t-j',j'}$, multiplexed and reordered by $\Pi_{t}^{\text{L}}$. The encoder in Fig.~\ref{Encoder}(a) corresponds to the special case $m=1$. 

In the DE recursion we now have to modify \eqref{eq:UpperUpdate3} to
\begin{equation*}
q_{\text{L}}^{(i-1)}=\epsilon \cdot\frac{\sum_{j=0}^{m} \sum_{k=0}^{m} \xs{L}{s}{i}{t+j-k}}{(m+1)^2} \ ,
\end{equation*}
and the a-posteriori erasure probability on the information bits at time $t$ and iteration $i$ (\ref{eq:appPCC}) becomes
 \[
p^{(i,t)}_{\rm a}=\epsilon\cdot \frac{\sum_{j=0}^{m} \sum_{k=0}^{m} \xs{U}{s}{i}{t+j}\xs{L}{s}{i}{t+k}}{(m+1)^2} \ .
\]

Likewise, for SC-SCCs the code sequence $\bs{v}_t^{\text{O}}$ of $\CO$ is divided randomly into the sequences  
$\tilde{\bs{v}}_{t,j}^{\text{O}}$, $j=0,\dots,m$. $\CI$ receives at time $t$ the sequences  $\tilde{\bs{v}}_{t-j,j}^{\text{O}}$ after passing a multiplexer and a permutation.  The encoder in Fig.~\ref{Encoder}(b) corresponds to the special case $m=1$. 

Equations \eqref{eq:InnerUpdate3} and \eqref{eq:OuterUpdate3} in the DE recursion are modified accordingly to
\begin{equation*}
q_{\text{O}}^{(i-1)}=\epsilon \cdot\frac{\sum_{j=0}^{m} \xs{O}{s}{i-1}{t-j}+\xs{O}{p}{i-1}{t-j}}{2(m+1)}
\end{equation*}
and
\begin{equation*}
q_{\text{I}}^{(i-1)}=\epsilon \cdot \frac{\sum_{j=0}^{m} \xs{I}{s}{i-1}{t+j}}{m+1} \ .
\end{equation*}
The a-posteriori erasure probability on the information bits at time $t$ after
$i$ iterations (\ref{eq:appSCC}) becomes
\[
p^{(i,t)}_{\rm a}=\epsilon  \cdot \xs{O}{s}{i}{t} \cdot \frac{\sum_{j=0}^{m} \xs{I}{s}{i}{t+j}}{m+1} \ .
\]

\section{Results and Discussion}
\label{secResults}

In this section, we give numerical results for some SC-TCs, using the DE described in Section~\ref{secDE} and~\ref{secMem}. In our examples we consider SC-TCs with identical rate-$1/2$, $4$-states component encoders. In particular, we consider component encoders with generator polynomials $(1,5/7)$ in octal notation. For notational simplicity, we denote the uncoupled PCC ensemble by $\CPCCa$ and the corresponding coupled ensemble by  $\CPCCSa$. For SC-SCCs, we denote by $\CSCCa$, and $\CSCCSa$ the uncoupled and coupled ensembles, respectively. Note that since the two component encoders are identical, $\fs{U}(x,y)=\fs{L}(x,y)$ and $\fp{U}(x,y)=\fp{L}(x,y)$ for SC-PCCs, and $\fs{I}(x,y)=\fs{O}(x,y)$ and $\fp{I}(x,y)=\fp{O}(x,y)$ for SC-SCCs. All presented thresholds correspond to the stationary case  $L \rightarrow \infty$, which lower bounds the thresholds for finite $L$.  For small $L$ the threshold can be considerably larger but at the expense of a higher rate loss.

In Table~\ref{Tab:PCC} we give the BP threshold for several SC-TCs and
coupling memory $m=1$ and $3$, denoted by $\epsilon_{\mathrm{SC}}^1$ and $\epsilon_{\mathrm{SC}}^3$ . We also report in the table the BP threshold ($\epsilon_{\mathrm{BP}}$) and the MAP threshold ($\epsilon_{\mathrm{MAP}}$) of the uncoupled ensembles. The MAP threshold was computed applying the area theorem \cite{Measson2009}. In all cases we observe an improvement of the BP threshold when coupling is applied. 
We remark that for $\CPCCSa$ the BP threshold of the uncoupled ensemble is already close to the MAP threshold, therefore the potential gain with coupling is limited. 
However, it is interesting to observe that the BP threshold of $\CPCCSa$ with $m=1$ is very close to $\epsilon_{\mathrm{MAP}}$, suggesting threshold saturation. 
%
The results for the ensemble $\CSCCSa$ are also given in Table~\ref{Tab:PCC} for coupling memory $m=1$ and $3$. We observe that the ensemble $\CSCCa$ has a poor BP threshold as compared to the MAP threshold. This is a well-known phenomenon for SCCs, for which the gap between the BP and the MAP threshold is large. A significant improvement is
obtained by applying coupling with $m=1$. However, there is still a gap between $\epsilon_{\mathrm{BP}}$ and $\epsilon_{\mathrm{MAP}}$, meaning that threshold saturation has not occurred. The BP threshold can be further improved by increasing the coupling memory to $m=3$. In this case the BP threshold is very close to the
MAP threshold, suggesting that threshold saturation occurs for large
enough coupling memory. This behavior is similar to the threshold
saturation phenomenon of SC-LDPC codes, which occurs for smoothing
parameter $w\rightarrow \infty$ \cite{Kudekar_ThresholdSaturation}.

\begin{table}[!t]
\caption{Thresholds for SC-TCs}
\vspace{-4ex}
\begin{center}\begin{tabular}{cccccc}
\toprule
Ensemble& Rate & $\epsilon_{\text{BP}}$ & $\epsilon_{\text{MAP}}$&
$\epsilon_{\text{SC}}^1$ &$\epsilon_{\text{SC}}^3$ \\
\otoprule
$\CPCCa$/$\CPCCSa$ & $1/3$ & 0.6428 & 0.6553 &0.6553 & 0.6553\\[0.5mm]
$\CSCCa$/$\CSCCSa$ & $1/4$ & 0.6896 & 0.7483 & 0.7378 & 0.7482\\[0.5mm]
\bottomrule
\end{tabular} \end{center}
\label{Tab:PCC} 
\end{table}

\begin{table}[!t]
\caption{Thresholds for punctured SC-TCs}
\vspace{-4ex}
\begin{center}\begin{tabular}{cccccc}
\toprule
Ensemble& Rate & $\epsilon_{\text{BP}}$ & $\epsilon_{\text{MAP}}$&$\epsilon_{\text{SC}}^1$ &$\epsilon_{\text{SC}}^3$ \\
\otoprule
$\CPCCa$/$\CPCCSa$ & $1/3$ & 0.6428 & 0.6553&0.6553 & 0.6553 \\[0.5mm]
$\CSCCa$/$\CSCCSa$ & $1/3$ & 0.6118 &0.6615 &0.6519 &0.6614\\[0.5mm]
\otoprule
$\CPCCa$/$\CPCCSa$ & $1/2$ &  0.4606 &0.4689& 0.4689 & 0.4689\\[0.5mm]
$\CSCCa$/$\CSCCSa$ & $1/2$ & 0.4010 &0.4973 &0.4773 & 0.4969\\[0.5mm]
\bottomrule
\end{tabular} \end{center}
\label{Tab:PCCpunctured} 
\vspace{-2ex}
\end{table}

In Table~\ref{Tab:PCCpunctured} we show the BP thresholds of punctured
SC-TCs, in order to compare SC-PCCs and SC-SCCs for a given code
rate. We consider $R=1/3$ and $R=1/2$, and coupling memory
$1$.\footnote{For $R=1/3$ the SC-PCC is not punctured.} For the SC-SCC
we used $\rho_1=1$ and $\rho_2=0.5$ for $R=1/3$ and $\rho_1=0.2$ and
$\rho_2=0.4$ for $R=1/2$. Again, in all cases an improvement of the BP
threshold is observed when coupling is applied. As expected, for a
given rate the PCC ensemble shows a better threshold than the SCC
ensemble. However, the improvement in the BP threshold due to coupling
for the latter is very significant. For $R=1/3$ and $m=1$ the BP threshold of
$\CSCCSa$ is very close to that of the (unpunctured) ensemble
$\CPCCSa$, while a large gap is observed for the uncoupled
ensembles. For $m=3$ $\CSCCSa$ achieves a better BP threshold than $\CPCCSa$. The result is even more remarkable for $R=1/2$. In this
case, while the uncoupled SCC ensemble shows a very poor threshold,
$\CSCCSa$ shows a superior threshold than $\CPCCSa$ already for $m=1$.

\section{ Conclusions}

In this paper we have introduced some block-wise spatially coupled
ensembles of parallel and serially concatenated convolutional codes
and performed a density evolution analysis on the BEC.  In all
considered cases spatial coupling results in an improvement
of the BP threshold and our numerical results suggest that threshold
saturation occurs if the coupling memory is chosen significantly
large. 
The threshold improvement is larger for the serial ensembles, which
are known to have poor BP thresholds without coupling but are stronger
regarding the distance spectrum. Puncturing the serial and parallel
ensembles to equal code rates, we observe that the threshold of the serial ensemble
can surpass the one of the parallel ensemble.


\bibliographystyle{IEEEbib}

\begin{thebibliography}{10}

\bibitem{JimenezLDPCCC}
{A. {Jim{\'e}nez} Feltstr{\"o}m} and {K.Sh. Zigangirov},
\newblock ``Periodic time-varying convolutional codes with low-density
  parity-check matrices,''
\newblock {\em {IEEE} Trans. Inf. Theory}, vol. 45, no. 5, pp. 2181--2190,
  Sept. 1999.

\bibitem{Kudekar_ThresholdSaturation}
S.~Kudekar, T.J. Richardson, and R.L. Urbanke,
\newblock ``Threshold saturation via spatial coupling: {W}hy convolutional
  {LDPC} ensembles perform so well over the {BEC},''
\newblock {\em {IEEE} Trans. Inf. Theory}, vol. 57, no. 2, pp. 803 --834, Feb.
  2011.

\bibitem{LentmaierTransITOct2010}
{M.~Lentmaier}, {A.~Sridharan}, {D.J.~Costello, Jr.}, and {K.Sh.~Zigangirov},
\newblock ``Iterative decoding threshold analysis for {LDPC} convolutional
  codes,''
\newblock {\em IEEE Trans.~Inf.~Theory}, vol. 56, no. 10, pp. 5274--5289, Oct.
  2010.

\bibitem{ldcc_2}
M.~Lentmaier, D.~Truhachev, and K.~Sh. Zigangirov,
\newblock ``To the theory of low density convolutional codes {II},''
\newblock {\em Problems of Information Transmission (Problemy Peredachi
  Informatsii)}, vol. 37, pp. 15--35, Oct.-Dec. 2001.

\bibitem{Hall2001}
E.K. Hall and S.G. Wilson,
\newblock ``Stream-oriented turbo codes,''
\newblock {\em IEEE Trans.~Inf.~Theory}, vol. 47, no. 5, pp. 1813--1831, Jul
  2001.

\bibitem{AAECC}
K.~Engdahl, M.~Lentmaier, and K.~Sh. Zigangirov,
\newblock ``On the theory of low-density convolutional codes,''
\newblock {\em Lecture Notes in Computer Science (AAECC-13)}, vol. 1719, pp.
  77--86, Springer-Verlag, New York 1999.

\bibitem{Huebner2008}
A.~Huebner, K.~{Sh.} Zigangirov, and D.~J. Costello, Jr.,
\newblock ``Laminated turbo codes: A new class of block-convolutional codes,''
\newblock {\em IEEE Trans.~Inf.~Theory}, vol. 54, no. 7, pp. 3024--3034, July
  2008.

\bibitem{ZhangBCC}
{W. Zhang}, {M. Lentmaier}, {K.Sh. Zigangirov}, and {D.J. Costello, Jr.},
\newblock ``Braided convolutional codes: a new class of turbo-like codes,''
\newblock {\em {IEEE} Trans. Inf. Theory}, vol. 56, no. 1, pp. 316--331, Jan.
  2010.

\bibitem{MoloudiISIT14}
S.~Moloudi and M.~Lentmaier,
\newblock ``Density evolution analysis of braided convolutional codes on the
  erasure channel,''
\newblock in {\em Proc. IEEE International Symposium on Information Theory},
  Honolulu, HI, USA, July 2014.

\bibitem{BerrouTC}
C.~Berrou, A.~Glavieux, and P.~Thitimajshima,
\newblock ``Near {Shannon} limit error-correcting coding and decoding:
  turbo-codes (1),''
\newblock in {\em Proc.\ {IEEE} International Conference on Communications},
  Geneva, Switzerland, May 1993, vol.~2, pp. 1064--1070.

\bibitem{Benedetto98Serial}
S.~Benedetto, D.~Divsalar, G.~Montorsi, and F.~Pollara,
\newblock ``Serial concatenation of interleaved codes: performance analysis,
  design, and iterative decoding,''
\newblock {\em {IEEE} Trans. Inf. Theory}, vol. 44, no. 3, pp. 909--926, May
  1998.

\bibitem{Kur03}
B.M. Kurkoski, P.H. Siegel, and J.K. Wolf,
\newblock ``Exact probability of erasure and a decoding algorithm for
  convolutional codes on the binary erasure channel,''
\newblock in {\em Proc.~IEEE Global Telecommunications Conference, 2003.
  GLOBECOM '03.}, Dec. 2003, vol.~3.

\bibitem{tenBrinkEXITConv}
J.~Shi and S.~ten Brink,
\newblock ``Exact {EXIT} functions for convolutional codes over the binary
  erasure channel,''
\newblock in {\em Proceedings of the 44th Allerton Conference on Communication,
  Control, and Computing}, Monticello, IL, USA, 2006.

\bibitem{AGiAa}
A.~Graell~i Amat, G.~Montorsi, and F.~Vatta,
\newblock ``Design and performance analysis of a new class of rate compatible
  serially concatenated convolutional codes,''
\newblock {\em {IEEE} Trans. Commun.}, vol. 57, no. 8, pp. 2280--2289, Aug
  2009.

\bibitem{AGiAb}
A.~Graell~i Amat, L.K. Rasmussen, and F.~Br\"annstr\"om,
\newblock ``Unifying analysis and design of rate-compatible concatenated
  codes,''
\newblock {\em {IEEE} Trans. Commun.}, vol. 59, no. 2, pp. 343--351, Feb 2011.

\bibitem{Measson2009}
C.~Measson, A.~Montanari, T.J. Richardson, and R.~Urbanke,
\newblock ``The generalized area theorem and some of its consequences,''
\newblock {\em IEEE Trans.~Inf.~Theory}, vol. 55, no. 11, pp. 4793--4821, Nov.
  2009.

\end{thebibliography}

\end{document}